\documentclass[
  aps,
  prx,
  twocolumn,
  superscriptaddress,
  longbibliography,
  nofootinbib,
  raggedbottom
]{revtex4-2}

\usepackage[T1]{fontenc}
\usepackage{placeins}
\usepackage{lmodern}
\usepackage{float}
\usepackage{microtype}
\usepackage{graphicx}
\usepackage{amsmath,amssymb,mathtools}
\usepackage{siunitx}
\usepackage{xcolor}
\usepackage{hyperref}
\hypersetup{colorlinks=true,linkcolor=blue,citecolor=blue,urlcolor=blue}

\begin{document}

\title{Contrastive local learning in large nonlinear analog circuits}

\author{Pragyan Pandey}
\affiliation{Department of Physics, University of California, Berkeley, California 94720, USA}

\author{Peter Wu}
\affiliation{Department of Electrical Engineering and Computer Sciences, University of California, Berkeley, California 94720, USA}

\author{Deema Alyahya}
\affiliation{Department of Physics, University of California, Berkeley, California 94720, USA}

\author{Eric Y. Ma}\email{eric.y.ma@berkeley.edu}
\affiliation{Department of Physics, University of California, Berkeley, California 94720, USA}
\affiliation{Department of Electrical Engineering and Computer Sciences, University of California, Berkeley, California 94720, USA}
\affiliation{Lawrence Berkeley National Laboratory, Berkeley, California 94720, USA}

\begin{abstract}
Machine learning in physical hardware promises to reduce training cost by colocating computation, memory and adaptation within device physics. Contrastive local learning networks (CLLNs) offer a compelling route: an electrical circuit relaxes to equilibrium in free and weakly clamped states, and each adaptive element updates from local voltage contrasts rather than backpropagated error signals. Yet whether nonlinear transistor implementations can scale beyond proof-of-concept circuits, and what design principles control that scaling, have remained unknown. Here we show that nonlinear analog circuits trained by contrastive local rules can learn larger machine-learning tasks and remain robust under hardware nonidealities. Using device-level circuit simulations, we systematically explore initialization, operating regime, loss-dependent clamping and training parameters, and scale CLLNs beyond ten thousand adaptive edges. The same local rule trains classifiers that reach the accuracy regime of conventional digital methods on Ionosphere and handwritten-digit benchmarks, and a network that autoregressively generates language over a 122-token vocabulary. These results establish the feasibility of local learning in large nonlinear analog circuits.

\end{abstract}

\maketitle

Modern digital machine learning is often constrained by data movement between physically separated memory and compute. Off-chip memory accesses can cost orders of magnitude more energy than a multiply--accumulate operation, while latency and bandwidth often dominate end-to-end performance. During training, this von Neumann bottleneck is especially severe: each optimization step requires repeated global passes over parameters, activations, gradients and optimizer state \cite{Horowitz2014, Sze2017DNN, Jouppi2017TPU, HennessyPatterson2019Turing}. These limitations have motivated a broad search for non--von Neumann systems that colocate computation, storage and adaptation within the same physical substrate, with the goal of faster and more energy-efficient learning \cite{Sebastian2020InMemory, McMahon2023NRPOpticalComputing, Momeni2025NaturePNN}.

Contrastive local learning networks (CLLNs) are a promising analog-learning platform at the intersection of neuromorphic computing and physical learning \cite{Stern2020Mechanical}. In a CLLN, an electrical network performs inference by relaxing under input voltages, while adaptive circuit elements store the learned parameters. Rather than using backpropagation through a digital model, CLLNs learn through coupled learning \cite{Stern2021PRX}: each edge updates from the local contrast between a free state and a weakly clamped state. Because updates are local and physically distributed, no centralized processor is needed to compute parameter updates during training. This makes CLLNs self-learning, parallel and robust to damage and local defects \cite{Dillavou2022PRApp}.

While early CLLN demonstrations with resistor edges were limited to learning linear functions, later implementations overcame this limitation using adaptive nMOS transistor edges. Unlike physical neural-network architectures that implement nonlinearity using separate activation elements \cite{Fu2024ONNReview,Shen2017Nanophotonic,Zuo2019AllOptical,Ashtiani2022PhotonicDNN,Xu2022ReconfigurableActivation}, employ stochastic sampling as a computational primitive \cite{Tuma2016StochasticNeurons,Camsari2017PBits,Dutta2022NeuralSampling,Choi2024PhotonicProbabilistic,Singh2024StochasticNanomagnets}, or depend on hybrid physical--digital training loops \cite{Yao2020MemristorCNN,Wright2022DeepPNN,Spall2022HybridTraining,Zheng2023DualAdaptive}, CLLNs utilize natively nonlinear transistors to combine nonlinearity, parameter storage and local adaptation within each edge. This makes transistor CLLNs a natural substrate for physical learning, and their compatibility with CMOS fabrication and peripheral circuitry suggests a path toward compact, high-density hardware realizations \cite{Markovic2020PhysicsNeuromorphic,Han2021CMOSNeuromorphic}.

However, nonlinear CLLNs have been demonstrated only in small proof-of-concept experiments, such as XOR and nonlinear regression, with a maximum of 32 transistor edges \cite{Dillavou2024PNAS}. Whether larger transistor CLLNs can solve higher-dimensional machine-learning tasks remains unexplored, and systematically investigating their design space in hardware is slow and labour-intensive. Simulation is therefore essential, but CLLN training requires two-phase circuit solves under changing boundary conditions, local gate-voltage updates and learning-level instrumentation, not natively supported by standard circuit simulators.

Here, we study moderate- and large-scale CLLNs using a SPICE-accurate framework that defines arbitrary topologies, solves free and clamped states, applies local contrastive gate-voltage updates and tracks learning across training. After validating the framework on canonical nonlinear tasks from Ref.~\cite{Dillavou2024PNAS}, we show that CLLNs can learn richer tasks, including image classification and generative language modelling, and remain robust under device mismatch and measurement noise. We further identify how initialization, operating regime, voltage clamping and training protocol shape learning, providing design principles for future hardware realizations.

\begin{figure*}[t]
    \centering
    \includegraphics[width=\textwidth]{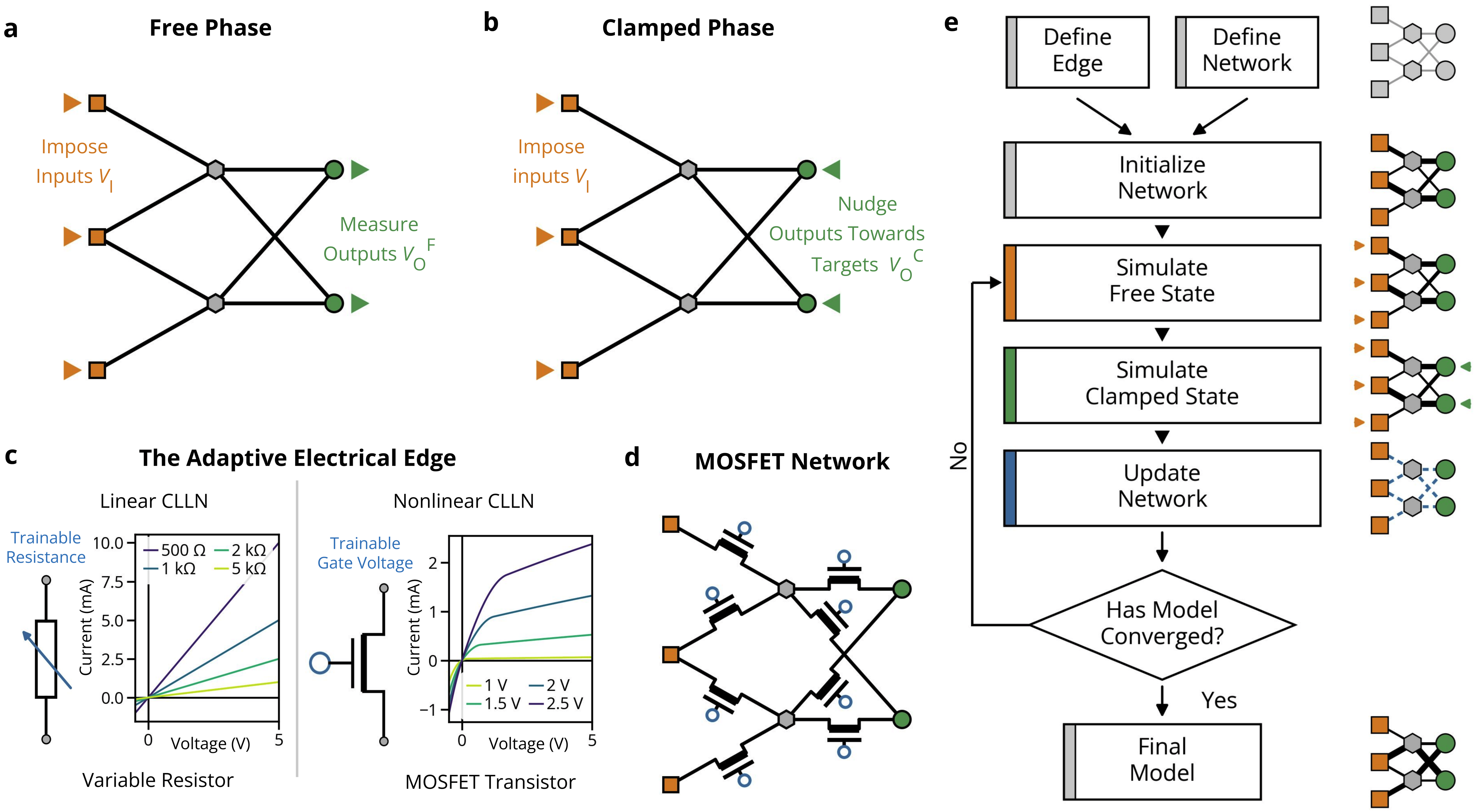}
    \caption{\textbf{Contrastive local learning in adaptive electrical networks.}
    \textbf{a}, Free phase: the input voltages $V_I$ are imposed and the free output voltages $V_O^F$ are measured. Orange squares, grey hexagons and green circles denote input, hidden and output nodes.
    \textbf{b}, Clamped phase: the same inputs are imposed while the outputs are nudged towards their targets, giving the clamped voltages $V_O^C$.
    \textbf{c}, Adaptive edges in linear and nonlinear contrastive local learning networks (CLLNs): a trainable resistance implements a linear edge; the trainable gate voltage $V_G$ of an nMOS transistor implements a nonlinear edge. Curves show current--voltage characteristics at the indicated parameter values.
    \textbf{d}, Example nonlinear CLLN with an nMOS transistor at every edge and $V_G$ as the trainable parameter.
    \textbf{e}, Simulation workflow: after the edge model and network are defined and initialized, free- and clamped-state solves and local updates repeat until convergence.}
    \label{fig:framework}
\end{figure*}

\subsection*{Contrastive local learning in electrical circuits}

A CLLN is represented as a graph in which nodes are voltage points and edges are adaptive electrical components (Fig.~\ref{fig:framework}a--d); inputs are imposed as voltages on input nodes and outputs are read at output nodes. The $N_I$ input and $N_O$ output nodes are fixed by the task, data representation and readout scheme, whereas the number of hidden nodes $N_H$ and the edge connectivity are design choices; this configuration is the \emph{circuit topology}. The network function $\mathcal{N}$ is determined by the topology and the $N_E$ edge learning degrees of freedom $\{G_1,G_2,\ldots,G_{N_E}\}$. The topology is thus analogous to the network architecture in conventional machine learning, while the edge degrees of freedom play the role of trainable weights:
\begin{equation}
\{V_{O1}, V_{O2}, \ldots, V_{O N_O}\}
=
\mathcal{N}\!\left(\{V_{I1}, V_{I2}, \ldots, V_{I N_I}\}\right).
\label{eq:network_function}
\end{equation}

Once a topology is chosen and initialized, training updates edges locally according to the coupled-learning rule \cite{Stern2021PRX}, which compares the circuit response in a free and a clamped phase. The rule is closely related to equilibrium propagation and contrastive Hebbian learning \cite{ScellierBengio2017EqProp, Movellan1991CHL}.  In the free phase (Fig.~\ref{fig:framework}a), the input voltages $V_I$ produce the natural output voltages $V_O^{F}$. In the clamped phase (Fig.~\ref{fig:framework}b), the same inputs are imposed while the outputs are constrained to voltages $V_O^{C}$ slightly closer to the desired solution than $V_O^{F}$. If the circuit relaxes by minimizing a physical quantity $P$ whose edge contributions can be evaluated locally, the parameter of edge $e$ evolves as
\begin{equation}
\dot{G}_e
=
-\gamma\,
\frac{\partial}{\partial G_e}
\left(P^{C}-P^{F}\right),
\label{eq:general_coupled_learning}
\end{equation}
where $P^F$ and $P^C$ are its free- and clamped-phase values and $\gamma$ is a learning-rate-like constant. These local edge updates progressively modify the circuit function $\mathcal{N}$ towards one that solves the task; the specific implementation of the update rule used in this work is described in Methods.

\section*{Results}

\subsection*{Efficient simulation of nonlinear CLLN circuits}
\label{sec:validation_xor_regression}

We developed \texttt{PyCLLN}, a framework for defining, training and analysing nonlinear CLLNs. Standard circuit simulators solve transistor networks numerically but offer no support for the training loop itself: arbitrary network construction, paired free- and clamped-state solves, loss-dependent clamps and local edge updates. \texttt{PyCLLN} integrates these operations while separating circuit topology, edge physics and learning objective into configurable components (Fig.~\ref{fig:framework}e). An adaptive edge is specified by a transistor SPICE model card with selectable body connection and gate-voltage reference; a \texttt{NetworkX} graph defines arbitrary arrangements of input, hidden and output nodes and is compiled into a parameterized SPICE netlist. Measurement noise and device-to-device mismatch can be introduced in a controlled manner.

Once compiled, the circuit is loaded through \texttt{NgSpiceShared} into a persistent simulator instance. The free-state solve is performed on this instance, the loss-dependent clamps are calculated, and the clamped-state solve reuses the same resident circuit; in noisy simulations, the free-phase and clamped-phase network copies coexist in one resident netlist (Methods). Gate voltages are then updated in place from local voltage contrasts. This persistence eliminates repeated netlist construction and simulator initialization, accelerating training by up to an order of magnitude (Supplementary Fig.~\ref{fig:supp_runtime}).

We validated the framework by reproducing nonlinear CLLN benchmarks of Ref.~\cite{Dillavou2024PNAS}. For XOR classification, the mean-squared error (MSE) fell by more than an order of magnitude as the circuit progressed from chance performance to correctly classifying all four inputs (Supplementary Fig.~\ref{fig:supp_validation}a,b). For nonlinear regression, the initially near-linear response evolved towards the target mapping while the MSE similarly decreased (Supplementary Fig.~\ref{fig:supp_validation}c,d). This validation of the simulation framework paves the way for the larger tasks explored below.

\subsection*{Scaling CLLNs for classification tasks}

We next scale transistor CLLNs beyond the 32-edge canonical circuits to radar classification and multiclass image recognition. These higher-dimensional tasks require substantially more adaptive edges and coordinated local updates than the canonical benchmarks, providing a stringent test of whether contrastive local learning continues to drive the network to a solution as it grows.

To train these larger networks, we introduce an output-clamping scheme derived from a margin-based hinge objective. Unlike the previously used MSE loss, which prescribes an absolute target voltage for every output, hinge loss requires only that the correct class output exceed its strongest competitor by a specified voltage margin. This removes the need to regress to a particular output vector and directly enforces the required class separation:
\begin{equation}
\mathcal{L}_{\mathrm{hinge}}
=
\max\!\left(0,\; m - \left(V_{Oy}^{(F)} - V_{Or}^{(F)}\right)\right),
\end{equation}
where \(m\) is the desired voltage margin, \(y\) is the true class, and \(r\) is the incorrect class with the highest free-phase output voltage. If the correct class already exceeds the top rival by at least \(m\), no clamped solve or edge update is performed. In practice this reduces training cost, especially once much of the training set satisfies the margin. When the margin is violated, only the correct class and the top rival are nudged in the clamped phase:
\begin{equation}
V_{Oy}^{C} = V_{Oy}^{F} + \frac{\delta}{2},
\qquad
V_{Or}^{C} = V_{Or}^{F} - \frac{\delta}{2},
\end{equation}
where \(\delta\) is the hinge clamp amplitude. This margin-gated, fixed-amplitude clamp is closely related to the overclamping protocol of Ref.~\cite{Dillavou2026Imperfection}, which was shown to improve robustness to noise and systematic bias in CLLNs.

\subsubsection*{Ionosphere classification}

\begin{figure*}[t]
    \centering
    \includegraphics[width=\textwidth]{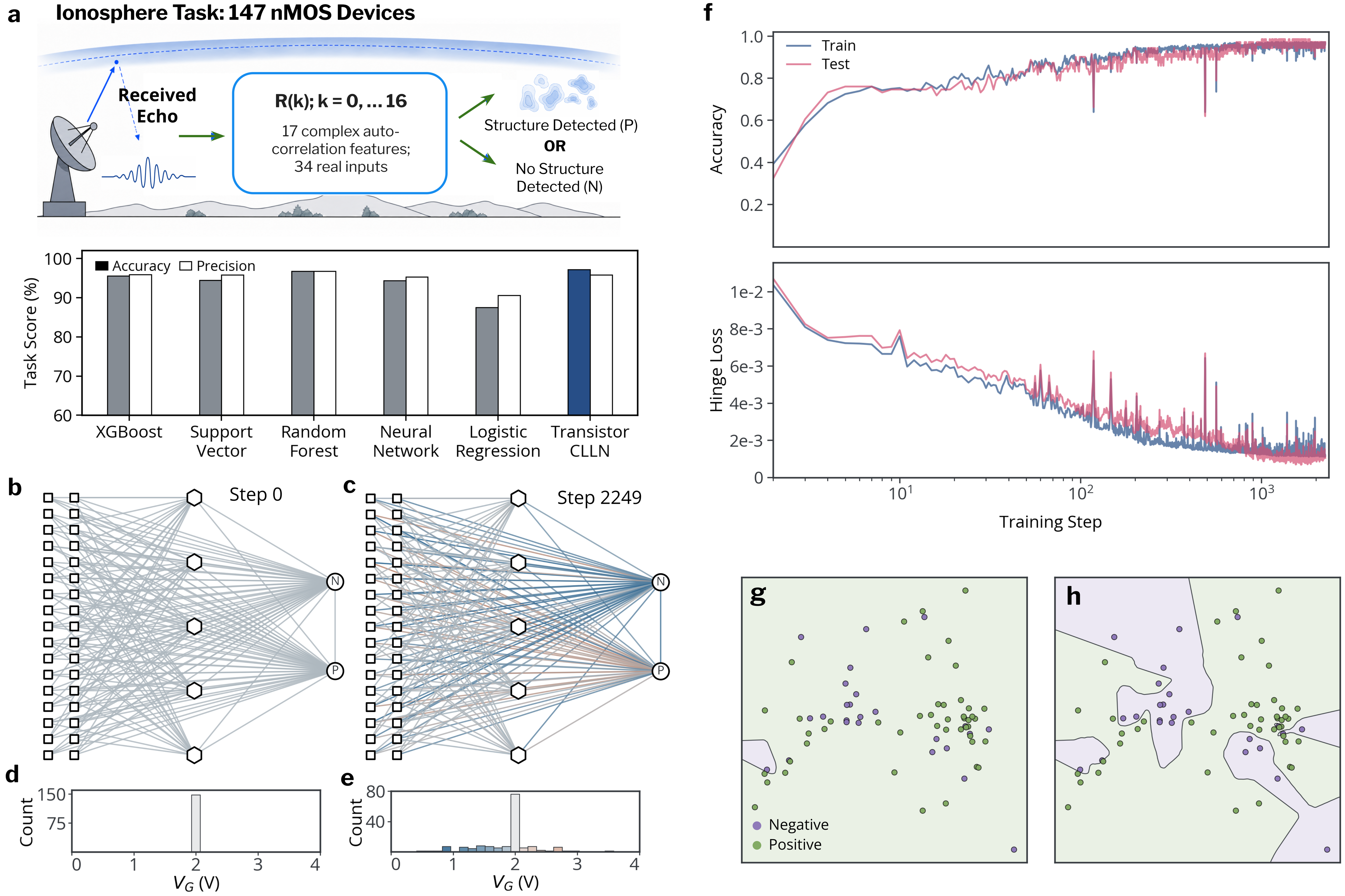}
    \caption{\textbf{Ionosphere classification with a transistor CLLN.}
    \textbf{a}, Each radar return is processed with an autocorrelation function at 17 pulse lags; the real and imaginary parts of the resulting 17 complex values give 34 input voltages. A 147-nMOS CLLN classifies each return as ionospheric structure (P) or no structure (N). Bars compare the final transistor-CLLN test accuracy and precision with published values for the indicated classifiers \cite{Sigillito1989UCI}.
    \textbf{b,c}, Gate-voltage states at initialization (\textbf{b}) and after training (\textbf{c}). Squares, hexagons and circles denote input, hidden and output nodes; edge colour encodes $V_G$.
    \textbf{d,e}, Corresponding $V_G$ distributions on the same colour scale.
    \textbf{f}, Training and test accuracy (top) and hinge loss (bottom) versus training step.
    \textbf{g,h}, Test samples and interpolated decision boundaries in a two-dimensional principal-component projection at initialization (\textbf{g}) and after training (\textbf{h}).}
    \label{fig:ionosphere}
\end{figure*}

The Ionosphere task is a binary classification problem comprising 351 returns collected by a high-frequency radar system at Goose Bay, Labrador \cite{Sigillito1989}. Each return is described by the real and imaginary components of the signal autocorrelation function at 17 pulse lags, yielding 34 continuous features. The task is to distinguish returns that contain evidence of structure in the ionosphere, denoted P, from those that do not, denoted N (Fig.~\ref{fig:ionosphere}a).

Of the 351 samples, we used 280 to train the network and 71 to evaluate its performance. The features were mapped to $[-0.8,0.8]$~V and applied directly to 34 input nodes. The circuit contained five hidden nodes and two output nodes, with each input connected to both outputs and to two hidden nodes, each hidden node connected to both outputs, and one lateral edge connecting the outputs. This topology comprised 147 adaptive nMOS devices, each with a floating body and the gate referenced to the drain. All gates were initialized to 2.0~V (Fig.~\ref{fig:ionosphere}b,d), and training used the hinge objective (training parameters in Methods). The predicted class was determined by the output with the higher free-phase voltage.

Notably, the gate-voltage distribution underwent only a slight reorganization as the circuit arrived at a solution (Fig.~\ref{fig:ionosphere}d,e). This limited reorganization also occurred selectively across the circuit (Fig.~\ref{fig:ionosphere}b,c). In terms of mean change, the input--hidden gates moved by only 4.7~mV, whereas the input--output, hidden--output and output--output gates moved by 546, 238 and 867~mV respectively, concentrating adaptation among the output-facing edges. A clear asymmetry also emerged between the two outputs: the mean gate voltage of the output-specific edges connected to N decreased from 2.00 to 1.45 V, whereas those connected to P increased to 2.13 V.

The CLLN, which initially assigned almost every sample to P, progressively learned a nonlinear decision function with distinct class regions (Fig.~\ref{fig:ionosphere}f--h). The trained network achieved a test accuracy of 97.2\%, correctly classifying 69 of 71 samples; this trained state is analysed in Fig.~\ref{fig:ionosphere}c,e,h. Published classifier results compiled with the dataset, including 87.5\% accuracy for linear logistic regression, provide conventional digital reference values (Fig.~\ref{fig:ionosphere}a) \cite{Sigillito1989UCI}. The transistor circuit therefore performs comparably to conventional digital classifiers on this radar task using only local contrastive updates.

We further trained three networks with device mismatch on all transistors and measurement noise on every voltage read during training and inference (Methods). These noisy realizations reached a mean test accuracy of 95.3\% $\pm$ 0.8\%, indicating that the classification remains robust to nonidealities (Supplementary Fig.~\ref{fig:supp_ionosphere_noise}).

\begin{figure*}[t]
    \centering
    \includegraphics[width=\textwidth]{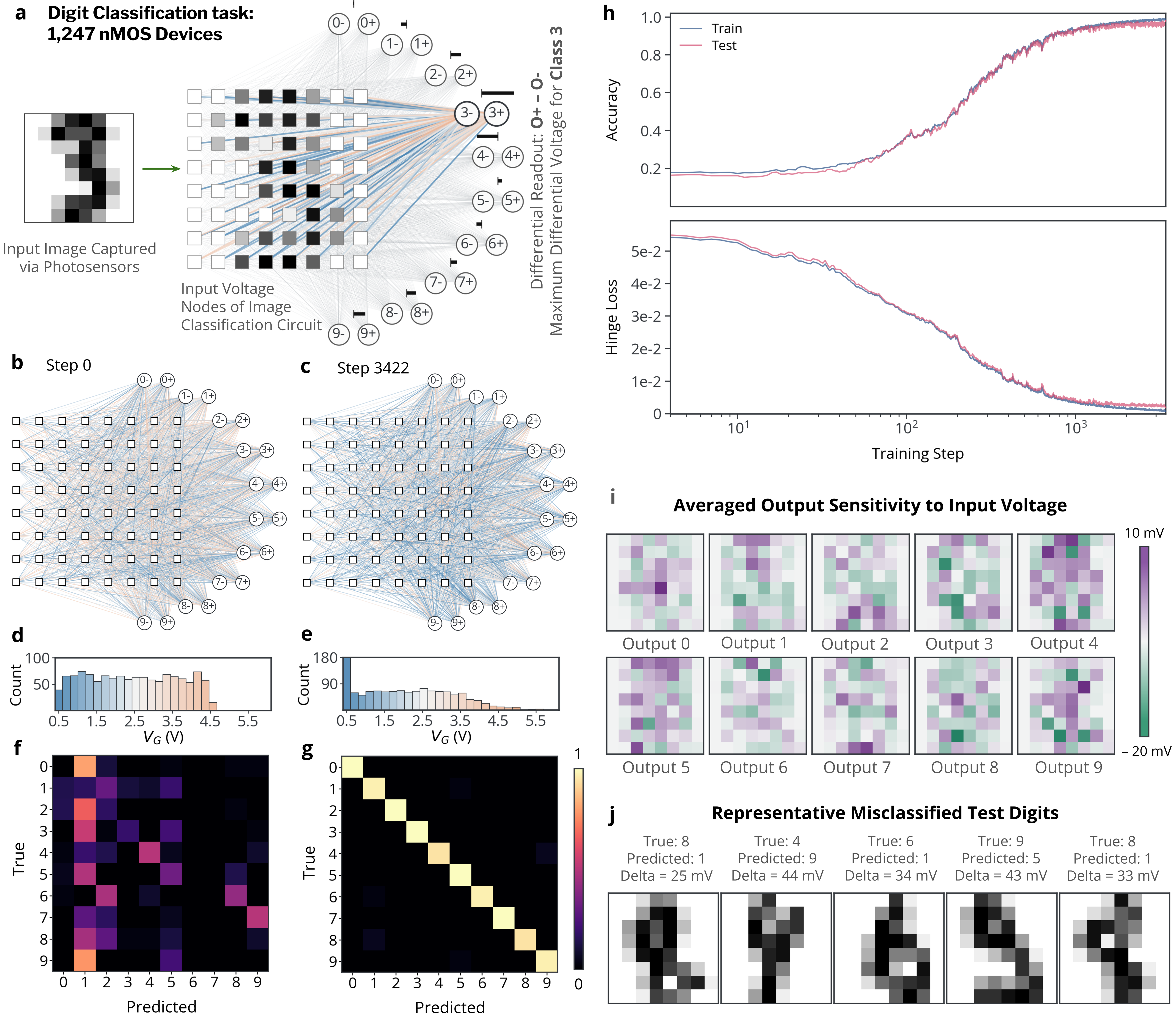}
    \caption{\textbf{Handwritten-digit classification with a transistor CLLN.}
    \textbf{a}, An $8\times8$-pixel image from the handwritten Digits dataset is converted to 64 input voltages and applied to a 1,247-nMOS CLLN with an output-node pair, $O_c^+$ and $O_c^-$, per class $c=0,\ldots,9$. Edges connected to the class-3 pair are coloured by gate voltage $V_G$; all other edges are grey. The bar above each pair shows its free-phase voltage difference, which is largest for class 3 for the input digit 3 shown, giving the correct prediction.
    \textbf{b,c}, Gate-voltage states at initialization (\textbf{b}) and after training (\textbf{c}); edge colour encodes $V_G$.
    \textbf{d,e}, Corresponding $V_G$ distributions on the same colour scale.
    \textbf{f,g}, Row-normalized test-set confusion matrices at initialization (\textbf{f}) and after training (\textbf{g}).
    \textbf{h}, Training and test accuracy (top) and hinge loss (bottom) versus training step.
    \textbf{i}, Input-weighted mean signed sensitivity of each differential readout to each input pixel over the test set: for output $c$ and pixel $p$, $S_{c,p}=\mathbb{E}_{\mathrm{test}}\!\left[x_p\,\partial D_c/\partial x_p\right]$, with $x_p$ the input voltage at pixel $p$.
    \textbf{j}, Representative misclassified test digits with true class $y$, predicted class $\hat{c}$ and output-margin error $\Delta=D_{\hat{c}}-D_y$.}
    \label{fig:digits}
\end{figure*}

\begin{figure*}[t]
    \centering
    \includegraphics[width=\textwidth]{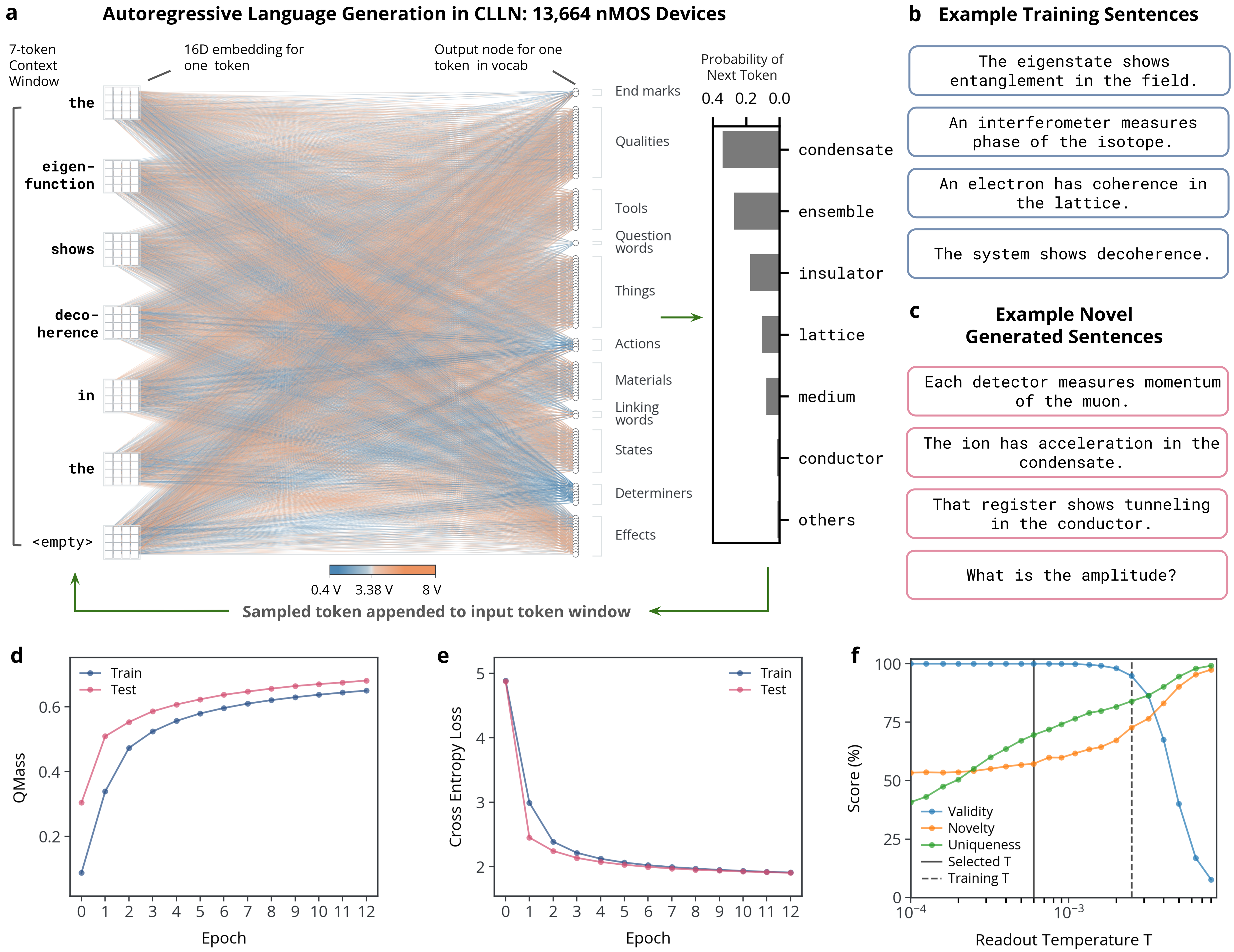}
    \caption{\textbf{Autoregressive language generation with a transistor CLLN.}
    \textbf{a}, A seven-token context window, encoded at 16 channels per token into 112 input voltages, connects through a 13,664-nMOS CLLN to 122 output nodes whose voltages define next-token logits over the 122-token vocabulary; edge colour encodes the trained $V_G$. A token sampled from the resulting distribution is appended to the context for autoregressive generation.
    \textbf{b}, Example training sentences from the scientific-language corpus.
    \textbf{c}, Example grammar-valid generated sentences absent from the training corpus.
    \textbf{d}, Training and test QMass, the mean probability mass assigned to grammar-valid next tokens, versus training epoch.
    \textbf{e}, Training and test next-token cross-entropy versus training epoch.
    \textbf{f}, Validity (fraction of grammar-valid sentences), novelty (fraction absent from the training corpus) and uniqueness (fraction of distinct sentences) of generated sentences versus readout temperature $T_{\mathrm{read}}$. Each point is the mean of five batches of 1,000 sentences. Solid and dashed lines mark $T_{\mathrm{read}}=0.6$~mV and the training temperature $T=2.5$~mV.}
    \label{fig:language}
\end{figure*}

\subsubsection*{Handwritten-digit classification}

We next used a transistor CLLN with 1,247 edges to classify handwritten digits. The scikit-learn Digits dataset comprises 1,797 $8\times8$-pixel greyscale images spanning ten digit classes \cite{Pedregosa2011}. We used 1,437 images for training and 360 for testing, with the 64 pixel intensities normalized to $[0,1]$~V and applied directly to 64 input nodes (Fig.~\ref{fig:digits}a). Each class $c$ was assigned a pair of output nodes, $O_c^+$ and $O_c^-$, whose voltage difference defined the free-phase class score,
$D_c=V_{O_c^+}-V_{O_c^-}$. The predicted digit was the class with the largest $D_c$. The adaptive nMOS devices were arranged exclusively as connections between the 64 input and 20 output nodes. Each transistor had its body tied to its drain and its gate voltage referenced to ground. The gate voltages were initialized randomly between 0.5 and 4.5~V (Fig.~\ref{fig:digits}b,d) and trained using the hinge objective (training parameters in Methods).

During training, hinge loss decreased smoothly while accuracy increased (Fig.~\ref{fig:digits}h). In contrast to the Ionosphere task, the gate-voltage distribution was markedly reshaped (Fig.~\ref{fig:digits}b--e). The gate voltages trended downward (mean 2.5~V to 2.123~V), and 225 of the 1,247 gates (18.0\%) moved below the nominal transistor threshold of 0.75~V, effectively turning off those edges. The device threshold therefore provides an intrinsic magnitude-pruning mechanism \cite{Han2015Pruning} reminiscent of tension-selective remodelling in biopolymer networks, where structurally insignificant filaments develop low tensions and are preferentially disassembled \cite{Cunha2024SelectivePruning,Wang2025Mechanosensitive}. The input-weighted sensitivity maps make the trained circuit interpretable: measured by small-signal analysis, they resolve each readout's score into per-pixel evidence (Fig.~\ref{fig:digits}i; unweighted maps in Supplementary Fig.~\ref{fig:supp_sensitivities}). The class-3 readout, for example, draws positive evidence from the top bar and lower-right tail of a 3, and penalizes left-side pixels that are empty in a 3 but inked in look-alikes such as 8, 5 and 2.

The untrained circuit assigned most test images to digit 1, whereas the trained circuit produced a nearly diagonal confusion matrix (Fig.~\ref{fig:digits}f,g). The trained network achieved a test accuracy of 98.06\%, correctly classifying 353 of 360 test images; the seven errors involved atypical digit forms, with representative examples in Fig.~\ref{fig:digits}j. For comparison, linear classifiers reach approximately 96\% on this task. Three networks retrained with device mismatch and measurement noise each reached a test accuracy of 96.9\% (Supplementary Fig.~\ref{fig:supp_digits_noise}). Thus, entirely local contrastive updates can coordinate more than one thousand nonlinear adaptive devices to achieve high-accuracy multiclass image recognition.

\subsection*{Autoregressive language generation with large CLLNs}

Finally, we scale CLLNs to autoregressive language generation with 13,664 adaptive nMOS devices. The task maps a context of up to seven tokens to a conditional probability distribution over the next token (Fig.~\ref{fig:language}a). Each token is represented by a frozen 16-dimensional SciBERT-derived embedding, so the seven-position context window produces $7\times16=112$ values that are scaled to $[0,0.45]$~V and applied as input voltages. These 112 inputs are fully connected to 122 output nodes representing the possible next tokens, producing 13,664 adaptive nMOS edges, the largest CLLN studied here. The free-phase voltages $V_{Oi}^{F}$ serve as the next-token logits.

To train this distributional output using contrastive local learning, we use a cross-entropy-based voltage clamp acting on all output nodes. First, the free-phase voltage outputs are converted to next-token probabilities:
\begin{equation}
p_i
=
\frac{
\exp\!\left(V_{Oi}^{F}/T\right)
}{
\sum_j \exp\!\left(V_{Oj}^{F}/T\right)
},
\label{eq:lm-softmax}
\end{equation}
where $T$ sets the sharpness of the softmax distribution; $T=2.5$~mV was used during training. For the one-hot target distribution $q_i$, the cross-entropy loss is
\begin{equation}
\mathcal{L}_{\mathrm{CE}}
=
-\sum_i q_i \log p_i.
\label{eq:lm-cross-entropy}
\end{equation}
The corresponding clamped outputs are
\begin{equation}
V_{Oi}^{C}
=
V_{Oi}^{F}
+
\delta\left(q_i-p_i\right),
\label{eq:lm-clamp}
\end{equation}
where $\delta$ is the clamp amplitude. The target output is therefore raised by $\delta(1-p_y)$, while each alternative output $i$ is lowered by $\delta p_i$, distributing the signal across the vocabulary according to the free-phase probabilities.

The network was trained on 5,762 sentences drawn from a controlled 122-token scientific vocabulary centred on quantum mechanics (examples in Fig.~\ref{fig:language}b). Each sentence generates multiple next-token prediction windows, yielding 52,604 training windows per epoch. The gate voltages were initialized randomly between 2.5 and 5.0~V and trained for 12 epochs (training parameters in Methods).

Over training, the CLLN progressively learned the grammatical constraints and lexical associations of the corpus. Cross-entropy loss decreased steadily on both the training and test sets, with the two curves approaching nearly identical values by the final epoch (Fig.~\ref{fig:language}e). Meanwhile, QMass, the mean probability mass assigned to grammar-valid next tokens, increased throughout training on both sets (Fig.~\ref{fig:language}d), concentrating probability on linguistically admissible continuations.

A more stringent test is whether these learned next-token relations remain useful when predictions are recursively composed into complete sequences. During generation, a sentence is constructed by successively sampling a token from the resulting distribution and appending it to the context before the next evaluation (Fig.~\ref{fig:language}a). This recursive process generated complete grammar-valid scientific sentences that had never appeared in the training corpus (Fig.~\ref{fig:language}c), demonstrating that the trained CLLN could compose learned token relationships into new sequences rather than simply reproduce memorized examples.

Generation is further tuned at readout without retraining the network. Increasing the readout temperature $T_{\mathrm{read}}$ broadens the sampled distribution, increasing novelty and uniqueness at the cost of admitting invalid continuations (Fig.~\ref{fig:language}f). At $T_{\mathrm{read}}=2.5$~mV, validity was 94.8\%, while novelty and uniqueness reached 72.7\% and 83.9\%, respectively. Lowering the temperature to $T_{\mathrm{read}}=0.6$~mV produced 100\% validity while retaining 57.2\% novelty and 69.5\% uniqueness. High validity therefore did not require collapse onto memorized training sequences. Generation also remained robust under device mismatch and measurement noise: three retrained realizations achieved a mean validity of 99.8\% (Supplementary Fig.~\ref{fig:supp_language_noise}).

\begin{figure*}[t]
    \centering
    \includegraphics[width=\textwidth]{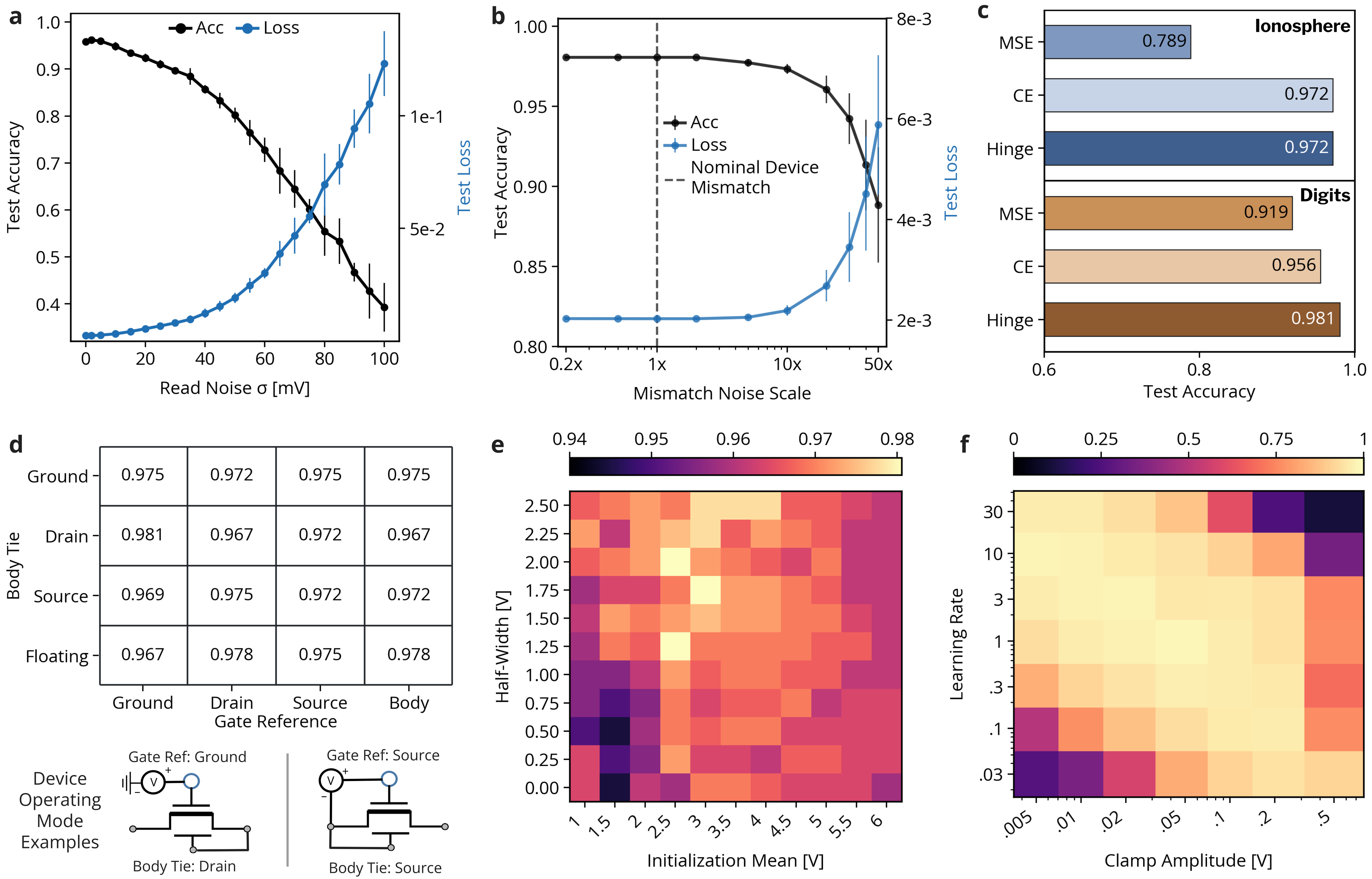}
    \caption{\textbf{Robustness and design sweeps for transistor CLLNs.}
    \textbf{a}, Digits test accuracy and hinge loss versus the standard deviation $\sigma$ of Gaussian voltage-read noise applied to every measurement during training and evaluation; each point is the mean $\pm$ s.d. of five independently trained networks.
    \textbf{b}, Test accuracy and hinge loss of the trained clean network under device mismatch at inference, each point the mean $\pm$ s.d. of five independent mismatch realizations. Threshold-voltage and strength variations scale together; the dashed line marks the ALD1106-based reference ($\sigma_{V_{\mathrm{th}}}=10\,\mathrm{mV}$, $\sigma_\beta/\beta=0.5\%$).
    \textbf{c}, Test accuracy on Ionosphere and Digits for mean-squared-error (MSE), cross-entropy (CE) and hinge objectives.
    \textbf{d}, Digits test accuracy across nMOS operating modes (body connection $\times$ gate-voltage reference); two example modes are shown below.
    \textbf{e}, Digits test accuracy versus the mean $\mu$ and half-width $h$ of the initial gate-voltage distribution $V_G\sim\mathcal{U}(\mu-h,\mu+h)$.
    \textbf{f}, Digits test accuracy versus learning rate $\gamma$ and clamp amplitude $\delta$. Panels \textbf{a} and \textbf{c--f} report the final test accuracy at the end of each run, using the fixed budgets specified in Methods; panel \textbf{b} evaluates the final trained network of the main clean run (Methods).}
    \label{fig:sweeps}
\end{figure*}

\subsection*{Noise robustness and hyperparameter considerations}

For physical implementation, contrastive local learning must tolerate hardware nonidealities while remaining effective across practical choices of circuit configuration and training parameters. We therefore examine these factors systematically, using the Digits network for the robustness and parameter sweeps, and both classification tasks for the learning-objective comparison (Fig.~\ref{fig:sweeps}).

We first examined robustness to voltage-read noise and device mismatch. Independent Gaussian noise was added to every measured voltage during training and evaluation, with five independently trained networks at each noise level (Fig.~\ref{fig:sweeps}a). The accuracy decreased progressively with increasing noise, remaining near 90\% up to $\sigma\approx30$~mV before deteriorating more strongly at larger amplitudes. To separately test the sensitivity of a learned solution to device variability, we evaluated the final trained network under static threshold-voltage and transistor-strength mismatch introduced at inference, jointly scaling both relative to a reference level of $\sigma_{V_{\mathrm{th}}}=10$~mV and $\sigma_\beta/\beta=0.5\%$ and averaging over five independent mismatch realizations at each level (Fig.~\ref{fig:sweeps}b). Little degradation occurred over more than an order-of-magnitude increase in mismatch, with accuracy declining and becoming more variable at much larger scales. Learning thus degrades progressively under measurement noise, whereas a trained solution tolerates far larger-than-reference device mismatch.

The learning objective and its output-clamping rule affected the accuracies achieved at the end of the training (Fig.~\ref{fig:sweeps}c). On both tasks, the hinge and cross-entropy clamps reached higher accuracy than the absolute-target MSE clamp and were comparable to each other, with equal accuracy on Ionosphere and hinge slightly ahead on Digits. We also varied the nMOS operating mode at each edge, namely its body connection and gate-voltage reference (Fig.~\ref{fig:sweeps}d). Despite the corresponding differences in transistor current--voltage response, all tested configurations reached comparably high Digits accuracy (96.7--98.1\%), with the body-drain, gate-ground configuration at the top of this narrow range.

Gate-voltage initialization and training parameters produced larger variations across the explored design space. Sweeping the initial distribution, $V_G\sim\mathcal{U}(\mu-h,\mu+h)$, revealed a broad high-performing region rather than a sharply defined optimum (Fig.~\ref{fig:sweeps}e). The highest accuracies generally favoured broad initial distributions, particularly for intermediate mean gate voltages ($\mu\approx2.5$--$3.5$~V) with broad half-widths ($h\approx1.25$--$2.5$~V), whereas fixed or narrowly distributed initializations performed less consistently, especially at lower mean gate voltages. By contrast, the joint sweep of learning rate $\gamma$ and clamp amplitude $\delta$ shaped performance sharply, with accuracies ranging from near chance-level to 98\% across the explored parameter space (Fig.~\ref{fig:sweeps}f). A broad diagonal region of high accuracy emerged, with smaller clamp amplitudes compensated by larger learning rates, whereas larger amplitudes favoured correspondingly smaller learning rates.

Rather than requiring a narrowly specified operating point, the networks thus remain robust across a sizeable region of the circuit and training design space.

\section*{Discussion and outlook}

This work establishes a SPICE-accurate framework for studying nonlinear CLLNs at scales difficult to access experimentally, and uses it to extend these systems from proof-of-concept circuits to networks of more than ten thousand adaptive transistors. Across all tasks, the same edge-local contrastive update trains circuits with different topologies, operating regimes and output objectives, without globally computed parameter updates. Beyond learning input--output mappings for classification, we show that CLLNs can represent context-dependent probability distributions whose outputs can be sampled recursively for autoregressive generation. Together, these results show that large nonlinear analog networks can learn qualitatively different forms of computation through the same distributed physical adaptation rule, while remaining robust to measurement noise and device mismatch.

Further scaling of CLLNs will benefit from continued study of circuit topology, device nonlinearity and update mechanisms. For circuit topology, the present results show that learning remains effective across different topologies and operating configurations. Broader exploration of depth and connectivity may help identify circuit designs that improve performance and reduce hardware complexity. Further work should also establish how different forms of device nonlinearity influence network computation and whether incorporating other adaptive elements, such as floating-gate MOS synapses \cite{Diorio1996SiliconSynapse}, ferroelectric field-effect transistor synapses \cite{Jerry2018FeFET}, metal-oxide memristors \cite{Prezioso2015Memristor} and Li-ion redox transistors \cite{Fuller2017LiIonSynapse}, can provide useful nonlinear responses for different tasks. Training also presently follows a stochastic gradient descent protocol with sample-wise updates, and therefore does not utilize the reduced update variance and computational parallelism available from mini-batch averaging \cite{Bottou2018Optimization}; a local physical implementation of such averaging remains an open challenge.

The robustness of CLLNs to measurement noise, device mismatch and variations in circuit and training parameters is encouraging for physical implementation. In particular, the persistence of high inference accuracy under substantial device mismatch suggests that learned solutions need not depend on the exact parameter values of the trained state and may tolerate moderate post-training drift, a known source of accuracy degradation in analog inference hardware \cite{Joshi2020PCM}. Such systems may be particularly attractive where signals are already analog, including in-sensor machine vision \cite{Mennel2020Nature,Zhou2019NeuromorphicVision}, adaptive biosensing \cite{Moin2021Biosensing,VanDoremaele2023Biosensor} and embedded sensorimotor control \cite{Krauhausen2021Sensorimotor}, allowing computation to remain in the analog domain. Together with the compatibility of transistor CLLNs with conventional electronic devices, these properties position nonlinear CLLNs as a promising platform for scalable, physically implemented learning and inference.

\section*{Methods}

\subsection*{Numerical circuit simulation}

\texttt{PyCLLN} is a Python-based framework that constructs parameterized CLLN netlists and coordinates their training through ngspice. Python manages circuit construction and the operations surrounding each circuit solve, including the application of voltage-encoded inputs, construction of loss-dependent output clamps and updating of gate voltages, while ngspice calculates the corresponding electrical equilibrium states. Each adaptive edge was instantiated as a parameterized SPICE subcircuit containing one transistor and a voltage source that set its gate voltage relative to a configurable reference terminal. The transistor body could be connected to ground, source or drain, or left floating, while the gate-voltage could be referenced to ground, source, drain or body. In this work, nMOS edges were modelled using the manufacturer-provided SPICE model for the ALD1106 device \cite{ALD1106Datasheet}.

Free states and clamped states requiring circuit relaxation were calculated as DC operating points in ngspice at $27\,^{\circ}\mathrm{C}$ using the KLU sparse solver; the complete solver configuration is included in the released netlists \cite{PyCLLN2026}. All simulations used ngspice 41 and were run on a single workstation with a 16-core AMD Ryzen 9 9950X processor and 192 GB of memory under Ubuntu 24.04.

\subsection*{Local contrastive update}

For each training example, the inputs were applied and the free state $F$ was solved before constructing the loss-dependent clamped state $C$. For edge $e$, with drain and source nodes denoted by $d(e)$ and $s(e)$, respectively, the voltage drop in phase $a\in\{F,C\}$ was
\begin{equation}
\Delta v_e^a=V_{d(e)}^a-V_{s(e)}^a.
\end{equation}
According to Ref. \cite{Dillavou2024PNAS}, the MOSFET gate-voltage update was then calculated as
\begin{equation}
\Delta V_{G,e}
=
-\gamma
\left[
\left(\Delta v_e^C\right)^2
-
\left(\Delta v_e^F\right)^2
\right],
\label{eq:methods-local-update}
\end{equation}
where $\gamma$ is the learning rate. The updated parameter was
\begin{equation}
V_{G,e}
\leftarrow
\operatorname{clip}
\left(
V_{G,e}+\Delta V_{G,e},
V_{G,e}^{\min},
V_{G,e}^{\max}
\right).
\end{equation}
The gate voltages were constrained to $[0.4,8.0]\,\mathrm{V}$. Because the update depends on the squared voltage drop, it is independent of the orientation assigned to an edge in the circuit graph. Training used sample-wise updates, with one gate-voltage update applied after each paired free and clamped solve.

\subsection*{Output clamping}

The learning objective is set through the construction of the clamped output state. For MSE, the free-state score $z_k^F$ was nudged towards a prescribed target $z_k^\star$ by
\begin{equation}
n_k=\eta\left(z_k^\star-z_k^F\right),
\label{eq:methods-mse-nudge}
\end{equation}
where $\eta$ is the nudge factor and $n_k$ is the score nudge. The hinge and cross-entropy clamping schemes and nudge calculations are defined in the main text. Constructing these two clamps requires comparisons across all output nodes: identifying the strongest rival for hinge, and normalizing the softmax for cross-entropy. These processes are confined to the readout nodes, whose count is far smaller than the number of edges that adapt through purely local voltage contrasts.

The resulting score nudges were translated into output-node voltages according to the readout scheme. For a single-ended output, $z_k=V_{O_k}$, the applied clamp was:
\begin{equation}
V_{O_k}^C=V_{O_k}^F+n_k.
\label{eq:methods-single-nudge}
\end{equation}
For a differential output, $z_k=D_k=V_{O_k^+}-V_{O_k^-}$, the nudge was divided evenly between the two output nodes:
\begin{equation}
V_{O_k^\pm}^C
=
V_{O_k^\pm}^F
\pm\frac{n_k}{2}.
\label{eq:methods-rail-nudge}
\end{equation}

\subsection*{Device mismatch and voltage-read noise}

Static device mismatch was introduced independently for each transistor by varying its threshold voltage and strength:
\begin{equation}
V_{\mathrm{TO},e}
=
V_{\mathrm{TO}}+\epsilon_{V,e},
\qquad
K_{\mathrm{P},e}
=
K_{\mathrm{P}}\left(1+\epsilon_{\beta,e}\right),
\label{eq:methods-mismatch}
\end{equation}
where $V_{\mathrm{TO},e}$ and $K_{\mathrm{P},e}$ are the threshold-voltage and process-transconductance parameters of transistor $e$, and $V_{\mathrm{TO}}$ and $K_{\mathrm{P}}$ are their nominal values. The threshold-voltage and relative-strength variations $\epsilon_{V,e}$ and $\epsilon_{\beta,e}$ were sampled independently from zero-mean Gaussian distributions with standard deviations $\sigma_{V_{\mathrm{th}}}$ and $\sigma_\beta/\beta$, respectively, where $\beta=K_{\mathrm{P}}W/L$ denotes transistor strength \cite{Pelgrom1989MOSMatching}. We defined the reference mismatch level as $\sigma_{V_{\mathrm{th}}}=10\,\mathrm{mV}$ and $\sigma_\beta/\beta=0.5\%$, based on ALD1106 specifications. In hardware CLLNs, the free and clamped states are realized simultaneously on two nominally identical copies of the network \cite{Dillavou2022PRApp,Dillavou2024PNAS}. Our noisy simulations follow this twin-network scheme: mismatch offsets were drawn independently for the free-phase and clamped-phase device sets, so that the local contrastive updates are also exposed to the systematic device disparity between the two circuit copies \cite{Dillavou2026Imperfection}. Each mismatch realization therefore comprises independent threshold-voltage and strength offsets for both device sets, sampled once and fixed throughout training and evaluation. Because evaluation and generation involve only free-state solves, all reported inference is performed on the free-phase device set.

Voltage-read noise was added independently to each measured voltage after every DC solve:
\begin{equation}
\widetilde{V}_i
=
V_i+\epsilon_i,
\label{eq:methods-read-noise}
\end{equation}
where $V_i$ is the noiseless SPICE voltage, $\widetilde{V}_i$ is the measured voltage and $\epsilon_i$ was sampled from a zero-mean Gaussian distribution with standard deviation $\sigma_{\mathrm{read}}$. Independent noise was sampled for every voltage measurement during the free and clamped phases, evaluation and generation. Read noise therefore perturbed the voltages used for learning and readout without altering the underlying electrical equilibrium. The reference read-noise level was $0.5\%$ of the maximum input-voltage magnitude for each task. Noisy runs otherwise used the corresponding clean data split, topology, initialization and learning parameters.

\subsection*{Validation tasks}

The implementation was validated using the canonical $4\times4$ periodic grid of 16 nodes and 32 adaptive nMOS edges from Ref.~\cite{Dillavou2024PNAS}, with the transistor bodies and gate-voltage sources referenced to ground. For XOR classification, two input nodes took values in $\{0,0.45\}\,\mathrm{V}$, two nodes were fixed at $0.11$ and $0.33\,\mathrm{V}$, and the differential output targets were 0 and $-87\,\mathrm{mV}$ for XOR-negative and XOR-positive examples, respectively. Gate voltages were initialized randomly in $[0.5,4.0]\,\mathrm{V}$ and trained for 10,000 epochs using $\eta=0.5$ and $\gamma=0.3$. The nonlinear-regression task used the same grid, with fixed-voltage nodes at 0 and $0.45\,\mathrm{V}$. All gate voltages were initialized to $3.0\,\mathrm{V}$ and trained for 25,000 epochs using $\eta=1$ and $\gamma=0.4$.

\subsection*{Classification tasks}

The 351 Ionosphere samples were divided using a seed-0 stratified 80:20 split into 280 training and 71 held-out evaluation samples. The supplied features, which lie in $[-1,1]$, were used without learned preprocessing and multiplied by $0.8\,\mathrm{V}$, producing input voltages in $[-0.8,0.8]\,\mathrm{V}$. The 1,797 Digits images were divided using the same procedure into 1,437 training and 360 held-out evaluation images. Pixel intensities were divided by 16 and applied directly as 64 input voltages in $[0,1]\,\mathrm{V}$. The exact circuit topology used for both tasks is released in \cite{PyCLLN2026}. In the Digits task, edge directionality was fixed such that transistor drains were connected to the inputs and sources were connected to the outputs; for the Ionosphere task, the edges connected to inputs have drains at inputs while hidden--output edges have drains connected to hidden nodes. The Ionosphere circuit was trained with the hinge objective using margin $m=10$~mV, clamp amplitude $\delta=10$~mV and learning rate $\gamma=4$; the Digits circuit used $m=20$~mV, $\delta=50$~mV and $\gamma=1.0$. The clean Ionosphere and Digits runs used fixed budgets of 50 and 15 epochs, respectively, and the reported accuracy is the final network state at the end of the budget. One training step denotes one applied gate-voltage update; for the hinge objective, samples that already satisfy the margin trigger no clamped solve and advance no step. Training follows a slower trajectory under noise for the Digits task (Supplementary Fig.~\ref{fig:supp_digits_noise}), so its noisy realizations were trained for a doubled budget of 30 epochs. The Ionosphere noisy realizations used the same 50-epoch budget as the clean run. Sensitivity maps (Fig.~\ref{fig:digits}i and Supplementary Fig.~\ref{fig:supp_sensitivities}) were computed by small-signal transfer analysis of the trained circuit at each test input's operating point; the input-weighted maps average $x_p\,\partial D_c/\partial x_p$ over the test set.

\subsection*{Language generation task}

The controlled quantum-mechanics corpus used 122 tokens and a beginning-of-sequence token, \texttt{<BOS>}, used only for padding. The complete token list and 6,998-sentence corpus are provided with the released code. The sentences were split into 5,762 training sentences and 1,236 test sentences. Each sentence was expanded into context--target windows, yielding 52,604 training windows per epoch; test metrics were evaluated on a fixed subset of 1,400 windows drawn from the test sentences.

Each non-\texttt{<BOS>} vocabulary token was encoded independently using the frozen \texttt{allenai/scibert\_scivocab\_uncased} embedding model \cite{Beltagy2019SciBERT}. The resulting 768-dimensional embeddings were $L_2$-normalized and standardized feature-wise across the vocabulary. Factor analysis then reduced each embedding to 16 dimensions for input to the CLLN, and each reduced component was independently scaled to $[0,0.45]\,\mathrm{V}$. The \texttt{<BOS>} embedding was set to zero.

The exact circuit topology used for the task is released in \cite{PyCLLN2026}; edge directionality was fixed such that transistor sources were connected to the inputs and drains were connected to the outputs. The network was trained with clamp amplitude $\delta=35$~mV and learning rate $\gamma=0.1$; the circuit architecture is described in the main text. Training continued for 12 epochs, and all reported quantities, including generation, use the final network state. For a context $\mathbf{c}$, QMass was defined as
\begin{equation}
Q(\mathbf{c})
=
\sum_{k\in\mathcal{A}(\mathbf{c})}
p(k\mid\mathbf{c}),
\label{eq:methods-qmass}
\end{equation}
where $\mathcal{A}(\mathbf{c})$ is the set of next tokens permitted by the complete grammar. QMass was averaged over the presented training windows and the fixed test subset.

Generation proceeded recursively until terminal punctuation was produced or 12 additional tokens had been sampled. A generated sequence was considered valid if it terminated and was valid under the grammar. Novelty was the fraction of generated sequences absent from the training corpus, and uniqueness was the fraction of distinct sequences. At each readout temperature, five independent batches of 1,000 sentences were generated.

\subsection*{Robustness and design sweeps}

Read-noise robustness was evaluated by independently training the main-text Digits circuit for 30 epochs, matching the noisy-training protocol, at 22 noise levels spanning $\sigma_{\mathrm{read}}=0$--$100\,\mathrm{mV}$; at each level, five networks were trained with independent read-noise streams and fixed initialization, and the reported value is the mean $\pm$ s.d. of their end-of-budget test accuracies. Device mismatch was disabled, and all other circuit and training settings were held fixed. Device-mismatch robustness was evaluated at inference using the clean network's final trained state, with voltage-read noise disabled. At each of 10 scales $s$ between 0.2 and 50, with $\sigma_{V_{\mathrm{th}}}=s(10\,\mathrm{mV})$ and $\sigma_\beta/\beta=s(0.5\%)$, five independent mismatch realizations were drawn and evaluated, and the reported value is their mean $\pm$ s.d. Threshold-voltage and strength mismatch were varied together, with the sampled transistor strength floored at zero.

Three additional sweeps examined device connections, gate-voltage initialization and learning parameters. The device-connection sweep tested all 16 combinations of body connection to ground, drain or source, or a floating body, and gate-voltage reference to ground, drain, source or body. For the initialization sweep, gate voltages were sampled as $V_G\sim\mathcal{U}(\mu-h,\mu+h)$, with $\mu$ varied from 1.0 to $6.0\,\mathrm{V}$ in $0.5\,\mathrm{V}$ increments and $h$ varied from 0 to $2.5\,\mathrm{V}$ in $0.25\,\mathrm{V}$ increments; sampled values were clipped to the permitted gate-voltage range. The learning-parameter sweep combined $\gamma\in\{0.03,0.1,0.3,1,3,10,30\}$ with $\delta\in\{5,10,20,50,100,200,500\}\,\mathrm{mV}$ at a fixed margin of $m=20\,\mathrm{mV}$. All configurations in these three sweeps were trained for 15 epochs using clean circuits, with all other settings not swept fixed to those of the main Digits experiment. The heatmaps report each configuration's final test accuracy at the end of its 15-epoch budget.

Learning objectives were compared for the Digits and Ionosphere tasks across the hinge, cross-entropy and MSE clamps (Fig.~\ref{fig:sweeps}c), with the clamped outputs constructed as defined in the Output clamping section. For each task the topology, device operating mode, training budget and data split were held at the main-experiment values, and each objective was run once at the main seed and reported at the end of the fixed budget (15 epochs for Digits, 50 for Ionosphere). The hinge cells reuse the main-experiment configuration of each task ($\gamma=4$, $\delta=10$~mV, margin $m=10$~mV for Ionosphere; $\gamma=1$, $\delta=50$~mV, margin $m=20$~mV for Digits), while the cross-entropy and MSE cells use per-objective training parameters. For Ionosphere, the cross-entropy run used $\gamma=2$, clamp amplitude $\delta=5$~mV and softmax temperature $T=5$~mV, and the MSE run used $\gamma=4$, nudge factor $\eta=0.02$ and a one-hot target of $0.5$~V, both retaining the main-experiment $2.0$~V gate initialization. For Digits, the cross-entropy run used $\gamma=1$, $\delta=100$~mV and $T=50$~mV with the main-experiment $\mathcal{U}(0.5,4.5)$~V initialization, and the MSE run used $\gamma=2$, nudge factor $\eta=0.1$ and a one-hot target of $0.5$~V with a wider $\mathcal{U}(2,8)$~V initialization.

\section*{Data availability}

The Ionosphere and Digits datasets analysed in this study are publicly available from the UCI Machine Learning Repository and scikit-learn, respectively. The scientific-language corpus generated in this study, together with the circuit topologies, device models, mismatch realizations, trained gate voltages and run metadata, is available with the \texttt{PyCLLN} source code \cite{PyCLLN2026}.

\section*{Code availability}

The PyCLLN source code and all scripts used to train and evaluate the circuits and reproduce the reported analyses are publicly available at Ref.~\cite{PyCLLN2026} under the MIT License. No access restrictions apply.

\section*{Funding}
The authors acknowledge the support of the George Lee Chair in Physics at the University of California, Berkeley.

\section*{Author contributions}
P.P. and E.Y.M. conceived the study. P.P., with contributions from P.W. and D.A., performed the simulations and analysis. E.Y.M. supervised the project. P.P. and E.Y.M. wrote the manuscript with input from all authors.

\section*{Competing interests}
The authors declare no competing interests.

\bibliographystyle{naturemag}
\bibliography{references}

\clearpage

\onecolumngrid

\setcounter{figure}{0}
\renewcommand{\thefigure}{\arabic{figure}}
\renewcommand{\theHfigure}{supp.\arabic{figure}}
\renewcommand{\figurename}{Supplementary Figure}
\setcounter{table}{0}
\renewcommand{\thetable}{S\arabic{table}}
\setcounter{equation}{0}
\renewcommand{\theequation}{S\arabic{equation}}

\begin{center}
    {\Large\bfseries Supplementary Information}\par
    \vspace{0.8em}
    {\large\bfseries Contrastive local learning in large nonlinear analog circuits}\par
    \vspace{1em}
    Pragyan Pandey,$^{1}$ Peter Wu,$^{2}$ Deema Alyahya,$^{1}$ and Eric Y. Ma$^{1,2,3,*}$\par
    \vspace{0.6em}
    {\small
    $^{1}$Department of Physics, University of California, Berkeley, California 94720, USA\par
    $^{2}$Department of Electrical Engineering and Computer Sciences, University of California, Berkeley, California 94720, USA\par
    $^{3}$Lawrence Berkeley National Laboratory, Berkeley, California 94720, USA\par
    $^{*}$Correspondence: eric.y.ma@berkeley.edu
    }
\end{center}

\vspace{1.5em}

\begin{figure}[H]
    \centering
    \includegraphics[
        width=0.9\textwidth,
        height=0.52\textheight,
        keepaspectratio,
        trim=180 0 180 0,
        clip
    ]{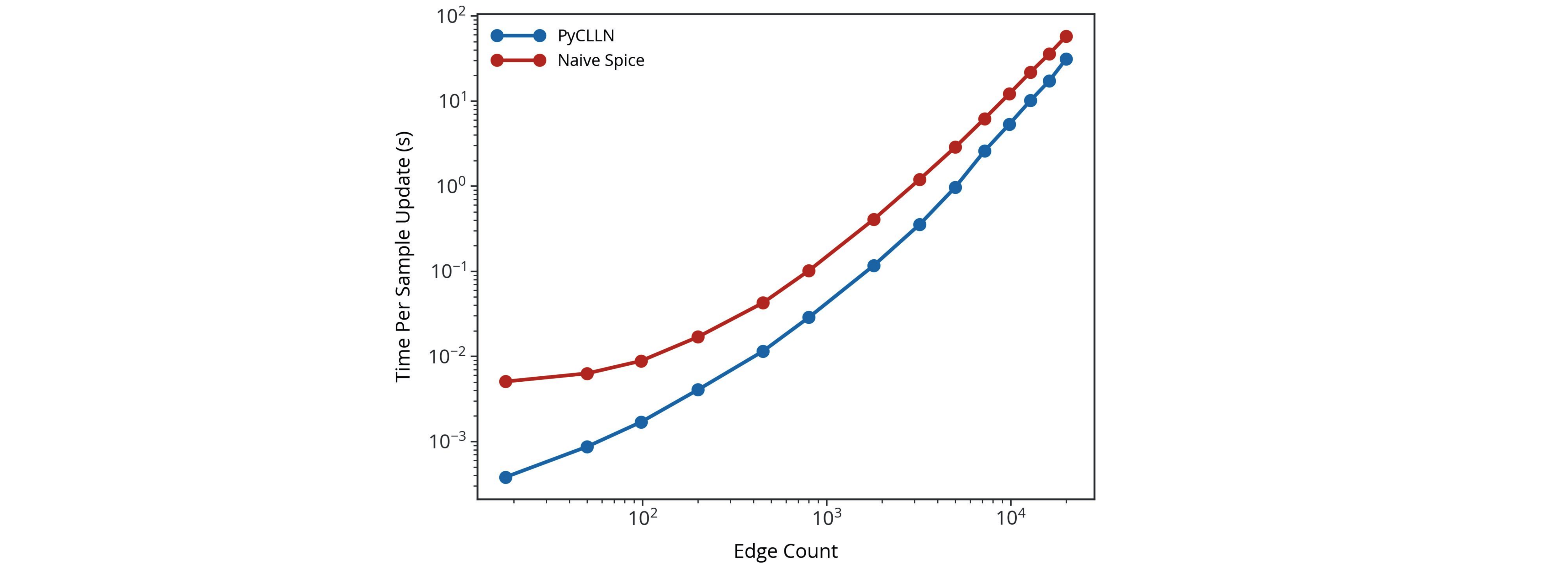}
    \caption{\textbf{Computational acceleration from persistent circuit simulation.}
    Wall-clock time per training-sample update versus the number of adaptive edges for the persistent-instance \texttt{PyCLLN} workflow and a naive workflow that constructs a new netlist and launches a fresh ngspice process for every solve; both use the KLU solver. Benchmarks use XOR-style $N\times N$ grid networks with $2N^2$ adaptive edges ($N=3$--$100$); each point is the mean per-sample update time over 50 epochs of the four-sample XOR truth table, measured on the workstation described in Methods (logarithmic axes).}
    \label{fig:supp_runtime}
\end{figure}

\clearpage

\begin{figure}[H]
    \centering
    \includegraphics[
        width=\textwidth,
        height=0.72\textheight,
        keepaspectratio
    ]{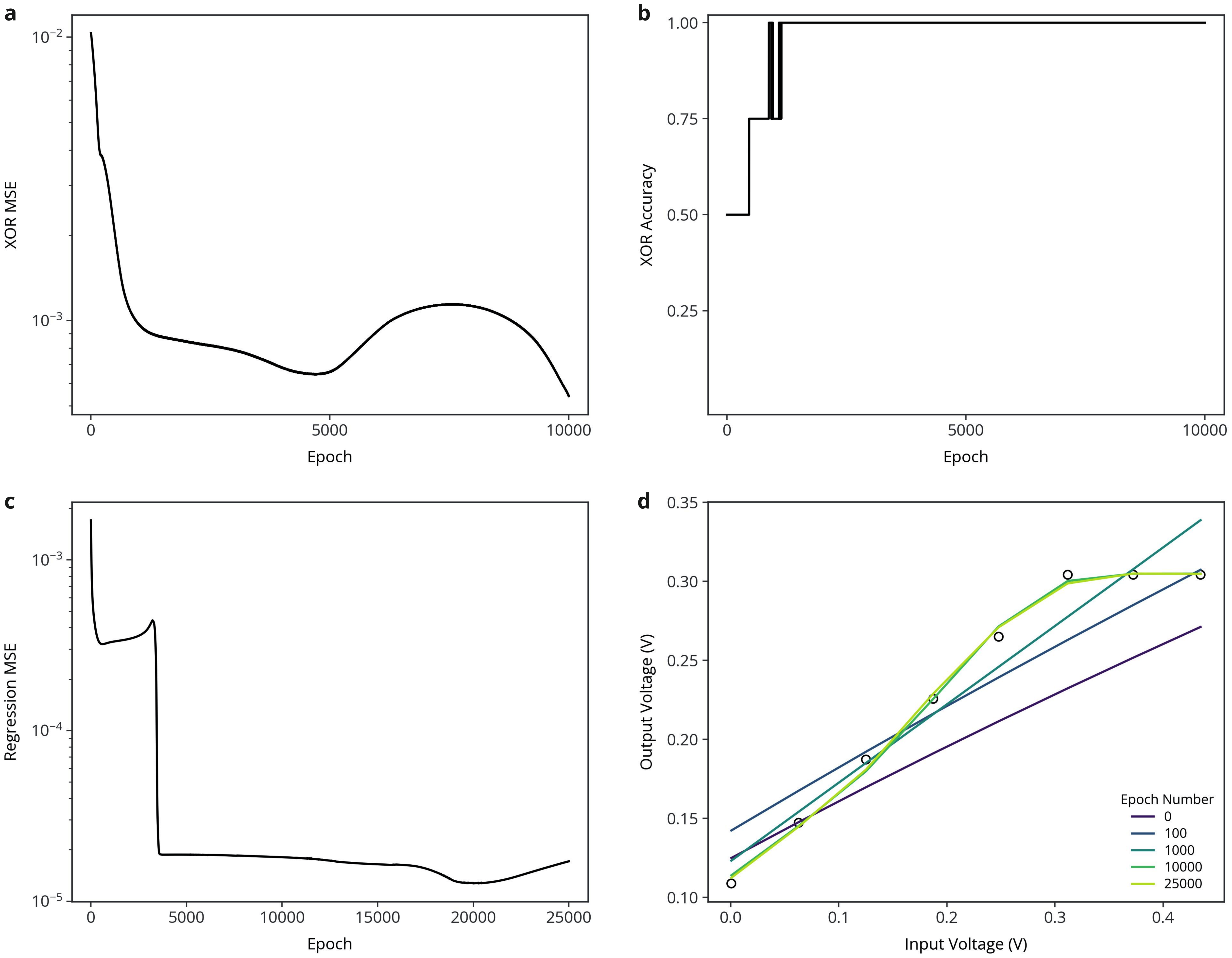}
    \caption{\textbf{Validation on canonical nonlinear CLLN benchmarks.}
    \textbf{a}, Mean-squared error (MSE) during 10,000 epochs of XOR training.
    \textbf{b}, Corresponding XOR classification accuracy.
    \textbf{c}, MSE during 25,000 epochs of nonlinear-regression training.
    \textbf{d}, Regression output versus input voltage at the indicated epochs; open circles mark the target mapping. Both tasks use the canonical $4\times4$ periodic circuit of 16 nodes and 32 adaptive nMOS edges.}
    \label{fig:supp_validation}
\end{figure}

\clearpage

\begin{figure}[H]
    \centering
    \includegraphics[
        width=\textwidth,
        height=0.72\textheight,
        keepaspectratio
    ]{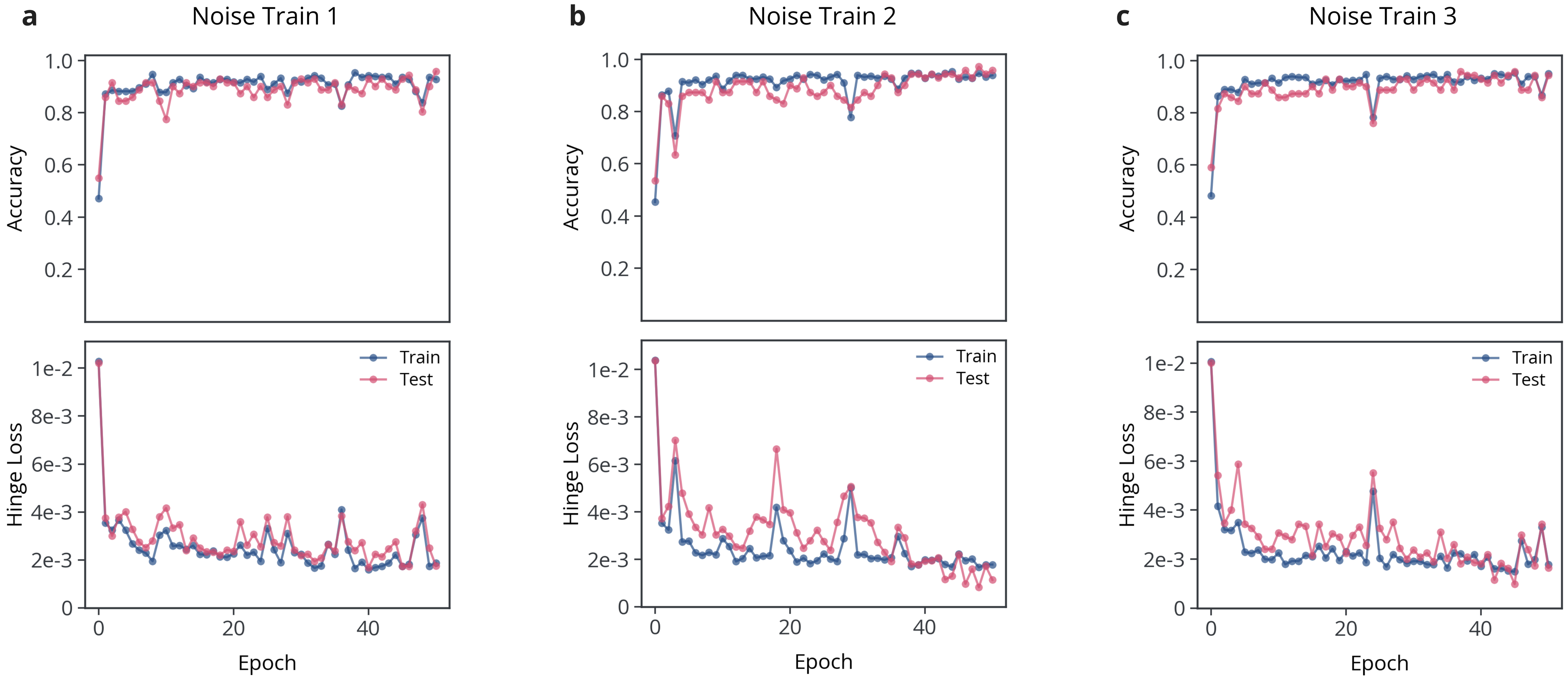}
    \caption{\textbf{Ionosphere training under device mismatch and voltage-read noise.}
    \textbf{a--c}, Training and test accuracy (top) and hinge loss (bottom) for three independent noisy realizations. Each network combines independently sampled static mismatch ($\sigma_{V_{\mathrm{th}}}=10\,\mathrm{mV}$, $\sigma_{\beta}/\beta=0.5\%$) with Gaussian read noise ($\sigma_{\mathrm{read}}=4.0\,\mathrm{mV}$) on every voltage measurement during training and evaluation, over a fixed 50-epoch budget. The reported accuracy is the final network state at the end of the budget; at that state the three realizations reached 95.8\%, 95.8\% and 94.4\% (68/71, 68/71 and 67/71).}
    \label{fig:supp_ionosphere_noise}
\end{figure}

\clearpage

\begin{figure}[H]
    \centering
    \includegraphics[
        width=\textwidth,
        height=0.62\textheight,
        keepaspectratio
    ]{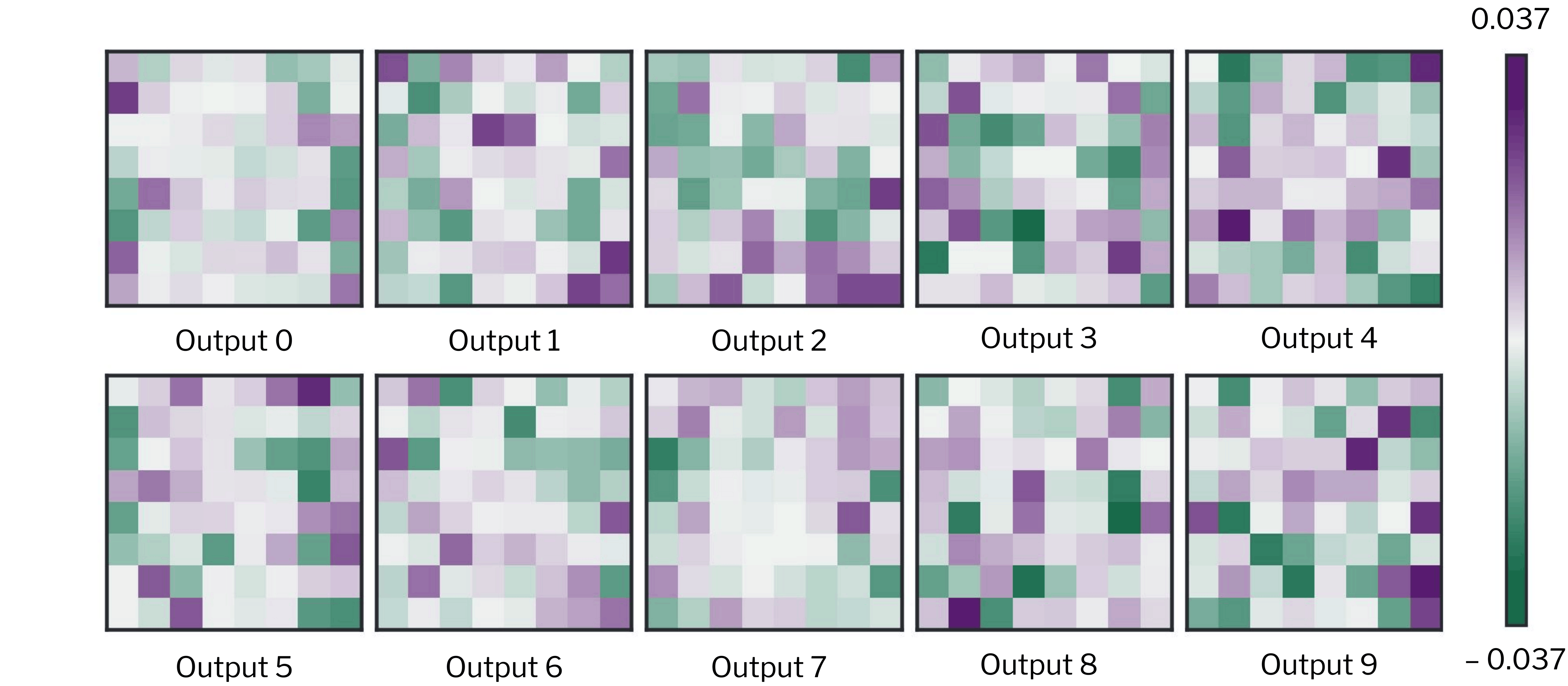}
    \caption{\textbf{Unweighted input sensitivities of the trained Digits CLLN.}
    For each digit-class output and input pixel, the map shows the test-set mean derivative of the differential class score (positive-rail minus negative-rail voltage) with respect to the input-pixel voltage, arranged in the $8\times8$ input-pixel layout. Positive values (purple) indicate that increasing the input voltage raises the class score; negative values (green) indicate that it lowers the score. Unlike the input-weighted sensitivities in Fig.~\ref{fig:digits}i, these values are not multiplied by the input voltage. Signed sensitivities arise from the differential rail pairs: each individual rail responds monotonically to the inputs because all transistor conductances are positive.}
    \label{fig:supp_sensitivities}
\end{figure}

\clearpage

\begin{figure}[H]
    \centering
    \includegraphics[
        width=\textwidth,
        height=0.72\textheight,
        keepaspectratio
    ]{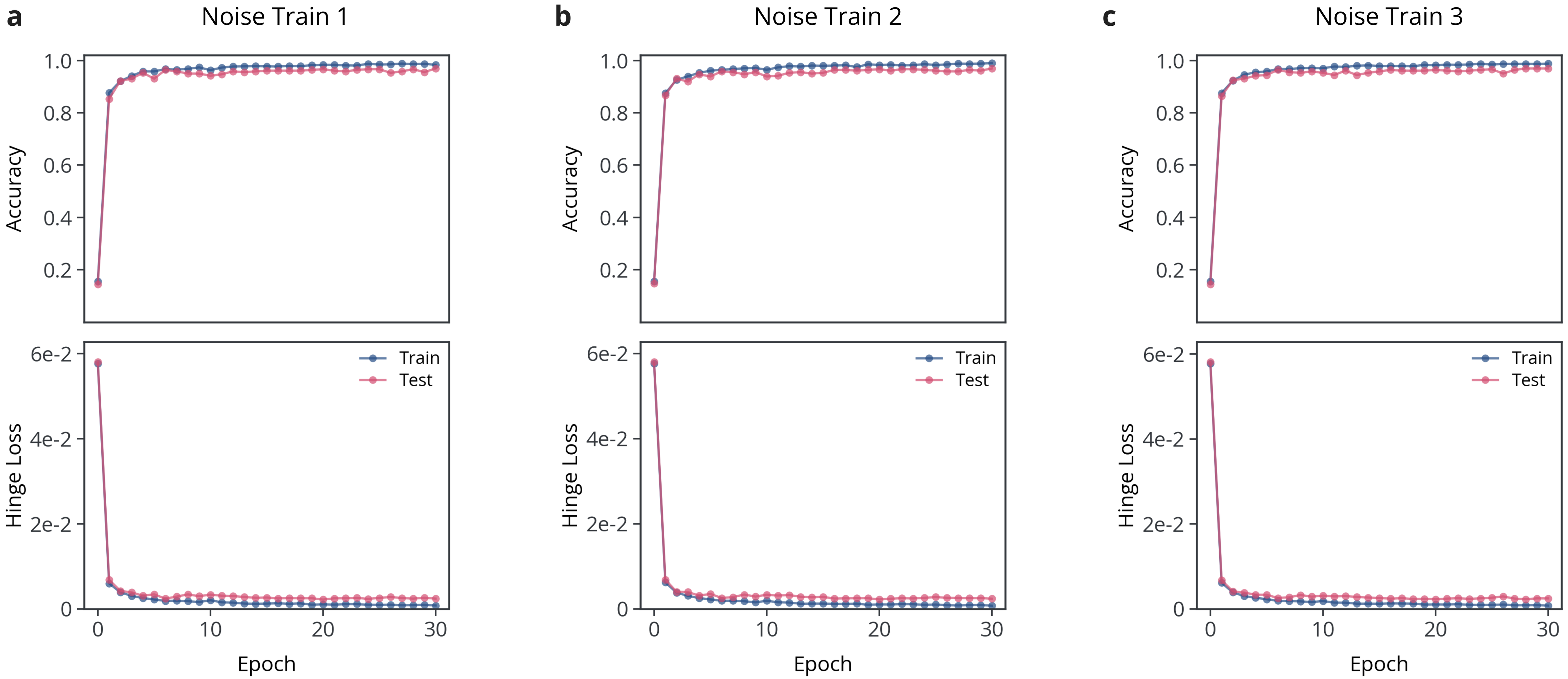}
    \caption{\textbf{Digits training under device mismatch and voltage-read noise.}
    \textbf{a--c}, Training and test accuracy (top) and hinge loss (bottom) for three independent noisy realizations. Each network combines independently sampled static mismatch ($\sigma_{V_{\mathrm{th}}}=10\,\mathrm{mV}$, $\sigma_{\beta}/\beta=0.5\%$) with Gaussian read noise ($\sigma_{\mathrm{read}}=5\,\mathrm{mV}$) on every voltage measurement during training and evaluation, over a fixed 30-epoch budget. At the end of the budget each of the three realizations reached a test accuracy of 96.9\%.}
    \label{fig:supp_digits_noise}
\end{figure}

\clearpage

\begin{figure}[H]
    \centering
    \includegraphics[
        width=\textwidth,
        height=0.72\textheight,
        keepaspectratio
    ]{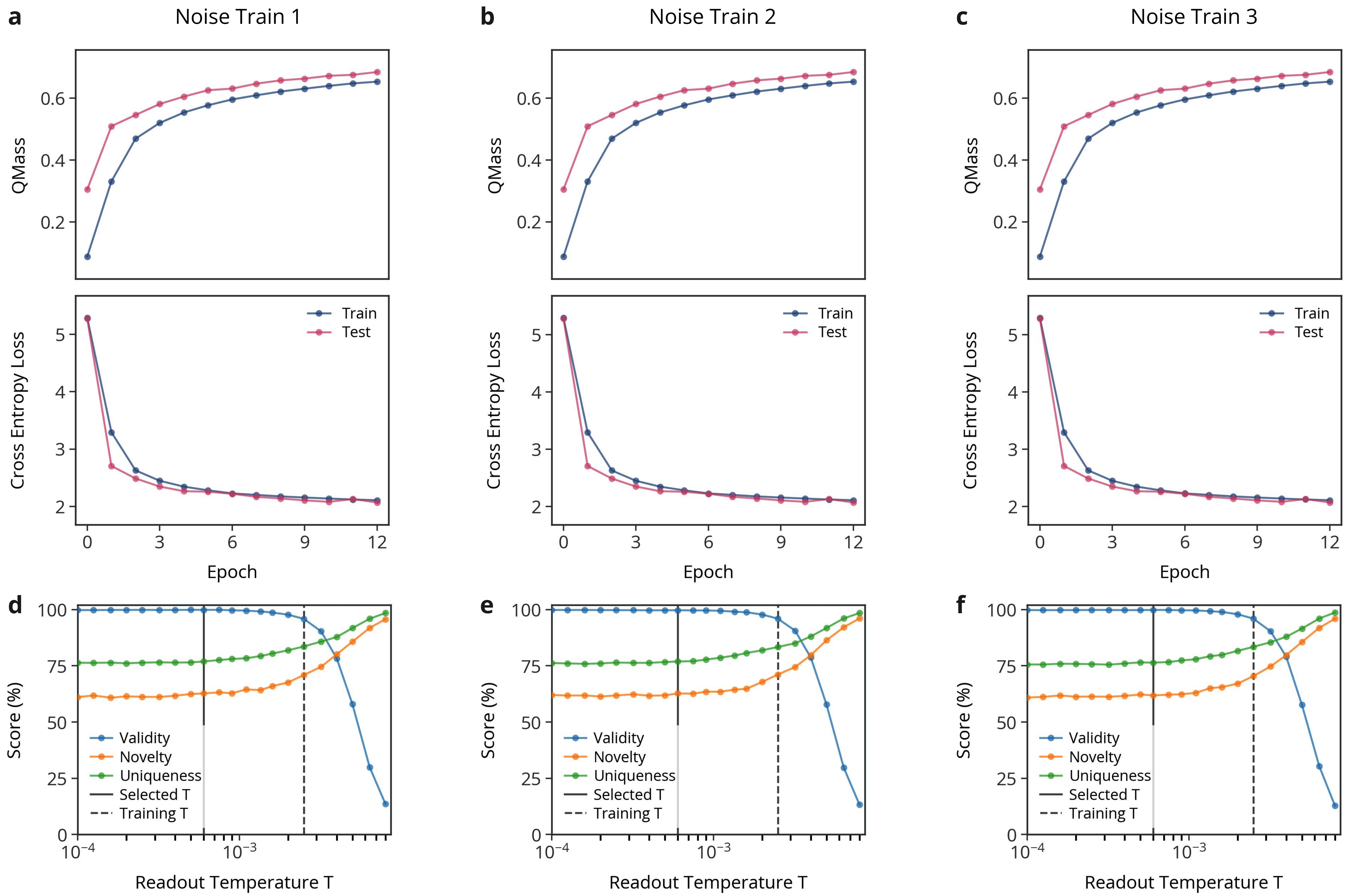}
    \caption{\textbf{Language-model training under device mismatch and voltage-read noise.}
    \textbf{a--c}, Training and test QMass (top; the probability mass assigned to grammar-valid next tokens) and cross-entropy loss (bottom) for three independent noisy realizations. Each network combines independently sampled static mismatch ($\sigma_{V_{\mathrm{th}}}=10\,\mathrm{mV}$, $\sigma_{\beta}/\beta=0.5\%$) with Gaussian read noise ($\sigma_{\mathrm{read}}=2.25\,\mathrm{mV}$) on every voltage measurement during training, evaluation and generation. The trajectories are nearly indistinguishable (final test QMass 0.6847, 0.6847 and 0.6846), although the trained gate states differ between realizations; the final epoch-12 state of each realization was used for generation.
    \textbf{d--f}, Validity, novelty and uniqueness of generated sentences versus readout temperature $T_{\mathrm{read}}$ for the corresponding realizations; each point is the mean of five batches of 1,000 sentences. Solid and dashed lines mark $T_{\mathrm{read}}=0.6$~mV and the training temperature $T=2.5$~mV. At $T_{\mathrm{read}}=0.6$~mV, validity was 99.86\%, 99.7\% and 99.82\% (mean 99.79\%).}
    \label{fig:supp_language_noise}
\end{figure}

\end{document}